\documentclass[aps,twocolumn,showpacs,superscriptaddress,preprintnumbers,floatfix,amsmath,amssymb]{revtex4-1}

\usepackage{amssymb}
\usepackage{graphicx}
\usepackage{dcolumn}
\usepackage{bm}
\usepackage{color}

\begin{document}

\title{Charge-density wave, superconductivity and $f$-electron valence instability in EuBiS$_{2}$F}

\author{Hui-Fei Zhai}
\thanks{These authors contribute equally to this work.}
\affiliation{Department of Physics, Zhejiang University, Hangzhou
310027, China}

\author{Zhang-Tu Tang}
\thanks{These authors contribute equally to this work.}
\affiliation{Department of Physics, Zhejiang University, Hangzhou
310027, China}

\author{Hao Jiang}
\thanks{These authors contribute equally to this work.}
\affiliation{Department of Physics, Zhejiang University, Hangzhou
310027, China}

\author{Kai Xu}
\affiliation{Department of Physics, Zhejiang University, Hangzhou
310027, China}

\author{Ke Zhang}
\affiliation{Department of Physics, Zhejiang University, Hangzhou
310027, China}

\author{Pan Zhang}
\affiliation{Department of Physics, Zhejiang University, Hangzhou
310027, China}

\author{Jin-Ke Bao}
\affiliation{Department of Physics, Zhejiang University, Hangzhou
310027, China}

\author{Yun-Lei Sun}
\affiliation{Department of Physics, Zhejiang University, Hangzhou
310027, China}

\author{Wen-He Jiao}
\affiliation{Department of Physics, Zhejiang University, Hangzhou
310027, China}

\author{I. Nowik}
\affiliation{Racah Institute of Physics, The Hebrew University,
Jerusalem 91904, Israel}

\author{I. Felner}
\affiliation{Racah Institute of Physics, The Hebrew University,
Jerusalem 91904, Israel}

\author{Yu-Ke Li}
\affiliation{Department of Physics, Hangzhou Normal University,
Hangzhou 310036, China}

\author{Xiao-Feng Xu}
\affiliation{Department of Physics, Hangzhou Normal University,
Hangzhou 310036, China}

\author{Qian Tao}
\affiliation{Department of Physics, Zhejiang University, Hangzhou
310027, China}

\author{Chun-Mu Feng}
\affiliation{Department of Physics, Zhejiang University, Hangzhou
310027, China}

\author{Zhu-An Xu}
\affiliation{Department of Physics, Zhejiang University, Hangzhou
310027, China} \affiliation{State Key Lab of Silicon Materials,
Zhejiang University, Hangzhou 310027, China} \affiliation{Center for
Correlated Matter, Zhejiang University, Hangzhou 310027, China}

\author{Guang-Han Cao} \email[Correspondence should be sent to: ]{ghcao@zju.edu.cn}
\affiliation{Department of Physics, Zhejiang University, Hangzhou
310027, China} \affiliation{State Key Lab of Silicon Materials,
Zhejiang University, Hangzhou 310027, China} \affiliation{Center for
Correlated Matter, Zhejiang University, Hangzhou 310027, China}

\begin{abstract}
Superconductivity (SC) and charge-density wave (CDW) are two
contrasting yet relevant collective electronic states which have
received sustained interest for decades. Here we report that, in a
layered europium bismuth sulfofluoride, EuBiS$_{2}$F, a CDW-like
transition occurs at 280 K, below which SC emerges at 0.3 K, without
any extrinsic doping. The Eu ions were found to exhibit an anomalously
temperature-independent mixed valence of about +2.2, associated
with the formation of CDW. The mixed valence of Eu gives rise to
self electron doping into the conduction bands mainly consisting of the in-plane Bi-6$p$ states, which in turn brings about the CDW and SC. In particular, the electronic specific-heat
coefficient is enhanced by $\sim$ 50 times, owing to the significant
hybridizations between Eu-4$f$ and Bi-6$p$ electrons, as verified by
band-structure calculations. Thus, EuBiS$_{2}$F manifests itself as
an unprecedented material that simultaneously accommodates SC, CDW
and $f$-electron valence instability.\\

\end{abstract}


\maketitle

\section{\label{sec:level1}Introduction}
Charge-density wave (CDW) and superconductivity (SC) are different
collective electronic orders, although both are associated with
Fermi surface instabilities owing dynamically to electron-phonon
interactions (for conventional BCS superconductors). CDW, usually occurring in low-dimensional materials,
generally shows periodic modulations of conduction electron density
and crystalline lattice in real space. In contrast, SC, appearing in
materials not limiting to low dimensionality, exhibits an intriguing
electronic ordering in momentum space due to condensation of Cooper
pairs, without any static lattice deformation. Basically they are
competing orders, nevertheless, coexistence of SC and CDW is
frequently observed in low-dimensional systems by various
experiments\cite{review,cdw,wilson}. In recent years, the relationship
between CDW and SC has become a hot topic in cuprate
high-temperature
superconductors\cite{review,wise,lawler,ghiringhelli,chang} as well
as 'conventional' superconductors that bear CDW
instability\cite{review,cdw,morosan}. Currently, CDW is also
considered as an intertwined electronic orders, not simply competing
with SC\cite{davis}.

Recently, SC was discovered in a quasi-two-dimensional (Q2D) bismuth
chalcogenide, LaO$_{1-x}$F$_{x}$BiS$_2$\cite{mizuguchi} (The
chemical formulae is preferably written as LaBiS$_2$O$_{1-x}$F$_{x}$
according to standard nomenclature\cite{nomenclature}). This new
class of materials consists of BiS$_2$ bilayers that are believed to
be responsible for SC. Band structure
calculations\cite{kuroki,wxg,li,yildirim} reveal that the undoped
parent compound LaBiS$_2$O belongs to a band insulator with an
energy gap of $\sim$ 0.8 eV\cite{wxg,li,yildirim}. The conduction
bands near Fermi level consist mainly of in-plane Bi-6$p$ orbitals.
Upon electron doping, these conduction bands are partially filled,
which leads to Q2D Fermi surface (FS) sheets. A minimal electronic
model including Bi- 6$p_x$ and 6$p_y$ orbitals was thus
constructed\cite{kuroki}. Interestingly, the resultant two bands
have a Q1D character with a double minimum dispersion, making FS
nesting possible. Possible CDW phases due to Q1D distortions of the
Bi and/or S atoms were proposed for $x\sim$ 0.5\cite{wxg,yildirim}.
Nevertheless, except for an inflection point in the temperature
dependence of resistivity in La$_{0.9}M_{0.1}$BiS$_2$O ($M$=Th, Ti,
Zr and Hf), which was speculated to be related to a CDW
effect\cite{yazici}, no more signatures in physical properties for a
CDW transition have been observed so far.

Here we report a series of evidences for a CDW transition at
$T_{\text{CDW}}\sim$ 280 K in an isostructural compound,
EuBiS$_{2}$F, synthesized for the first time. Unlike other parent
compounds such as LaBiS$_2$O\cite{LBSF} and SrBiS$_2$F\cite{lei}
that are undoped insulators, surprisingly, EuBiS$_{2}$F itself is
metallic, and moreover it exhibits SC below 0.3 K. By various
experimental approaches, we demonstrate that EuBiS$_{2}$F is
actually self doped due to partial electron transfer from the Eu
ions to the BiS$_2$ bilayers. The Eu ions exhibit an anomalously
temperature-independent mixed valence of about +2.2. The electronic
specific-heat coefficient extracted from the experimental data is as
large as 73 mJ K$^{-2}$ mol$^{-1}$, $\sim$ 50 times larger than
those of its analogues, suggesting significant hybridizations
between Eu-4$f$ and Bi-6$p$ electrons. Therefore, to our knowledge,
EuBiS$_{2}$F represents the first material that simultaneously bears
SC, CDW and $f$-electron valence instability.

\section{\label{sec:level2}Experimental Methods}

\paragraph{Sample's synthesis} The EuBiS$_{2}$F polycrystalline sample
was synthesized by a solid-state reaction in sealed evacuated quartz
tubes. All the starting materials were bought from Alfa Aesar. The
stoichiometric mixtures of EuS [presynthesized by reacting Eu
(99.9\%) and S (99.9995\%) pieces in sealed evacuated quartz tubes
at 1073 K for 20 h], EuF$_2$ (99.9\%) and Bi$_2$S$_3$ (99.995\%)
powders, loaded in an evacuated quartz ampule, were heated in a
muffle furnace to 1053 K for 20 h. The reacted mixtures were ground
for homogenization in an agate mortar, pressed into pellets, and
sintered at 1053 K for another 20 h. This process was repeated until
nearly single-phase sample was obtained. An argon-filled glove box
was employed for the operations above to avoid the contamination of
water and oxygen as far as possible.

\paragraph{X-ray diffractions and crystal structure} Powder x-ray
diffraction (XRD) was carried out at room temperature and at low temperatures
down to 13 K on a PANalytical x-ray diffractometer (Model EMPYREAN)
with a monochromatic CuK$_{\alpha1}$ radiation. The lattice
parameters were precisely determined using Si powders as the
internal standard reference material. The crystal structure was
refined based on the CeBiS$_2$O-type structure model\cite{ceolin} by
a Rietveld analysis using the code Rietan-2000\cite{rietan}. With
the exact lattice parameters, all the structural refinements were easily
convergent. The resultant weighted reliable factor $R_{\text{wp}}$
is 7.1$-$9.0\%, and the 'goodness of fit' parameter $S$ is
1.1$-$1.5, indicating good reliability for the refined crystal structure.

\paragraph{Physical property measurements} The electrical resistivity
was measured with a standard four-electrode method on a Quantum
Design PPMS-9. The as-prepared EuBiS$_{2}$F pellet was cut into a thin bar with
a dimension of 2.0$\times$1.6$\times$0.6 mm$^3$, and gold wires
($\phi$=30 $\mu$m) were attached onto the sample's surface with
silver paint. The size of the contact pads produces a total
uncertainty in the absolute values of resistivity of ¡À15\%. A
dilution refrigerator was employed for measuring the resistivity at
ultra low temperatures down to 20 mK with an electric current of 20
$\mu$A. Hall coefficient was measured by permutating the voltage and
current electrodes\cite{sample} under a magnetic field of 60 kOe,
and the results were checked by a standard field-sweep measurement.
The heat capacity was measured using a relaxation method on a
PPMS-9. The dc magnetization was measured on a Quantum Design MPMS-5
equipment. Signals from the sample holder were carefully removed.

\paragraph{M\"{o}ssbauer spectroscopy} M\"{o}ssbauer studies on $^{151}$Eu at various temperature (up to
388 K), were performed by using a conventional constant acceleration
drive and $\sim$50 mCi $^{151}$Sm$_2$O$_3$ source. All spectra
obtained were analyzed in terms of least square fit procedures to
theoretical expected spectra. The experimental spectra were analyzed
by two Lorentzian lines from which values for the isomer shift
($S$) and the spectral area of the resonance absorption lines were
derived. The analysis considered also the exact shape of the source
emission line\cite{felner}. The velocity calibration was performed
with an $\alpha$-iron foil at room temperature and the reported $S$
values are relative to Eu$_2$O$_3$ at room temperature.

\paragraph{Electronic structure calculations} We carried out
electronic structure calculations using the Vienna Ab-initio
Simulation Package (VASP)\cite{vasp}. The experimental crystal
structure parameters at 15 K were employed for the calculations. The
strong Coulomb repulsion of the Eu-4$f$ electrons was included using
local spin density approximation plus a $U$ parameter (LSDA+$U$).
The plane-wave basis energy cutoff was set at 540 eV.

\section{\label{sec:level3}Results and discussions}

\subsection{\label{subsec:level1}Crystal structure and bond valence sum of Eu}
\begin{figure*}
\includegraphics[width=16cm]{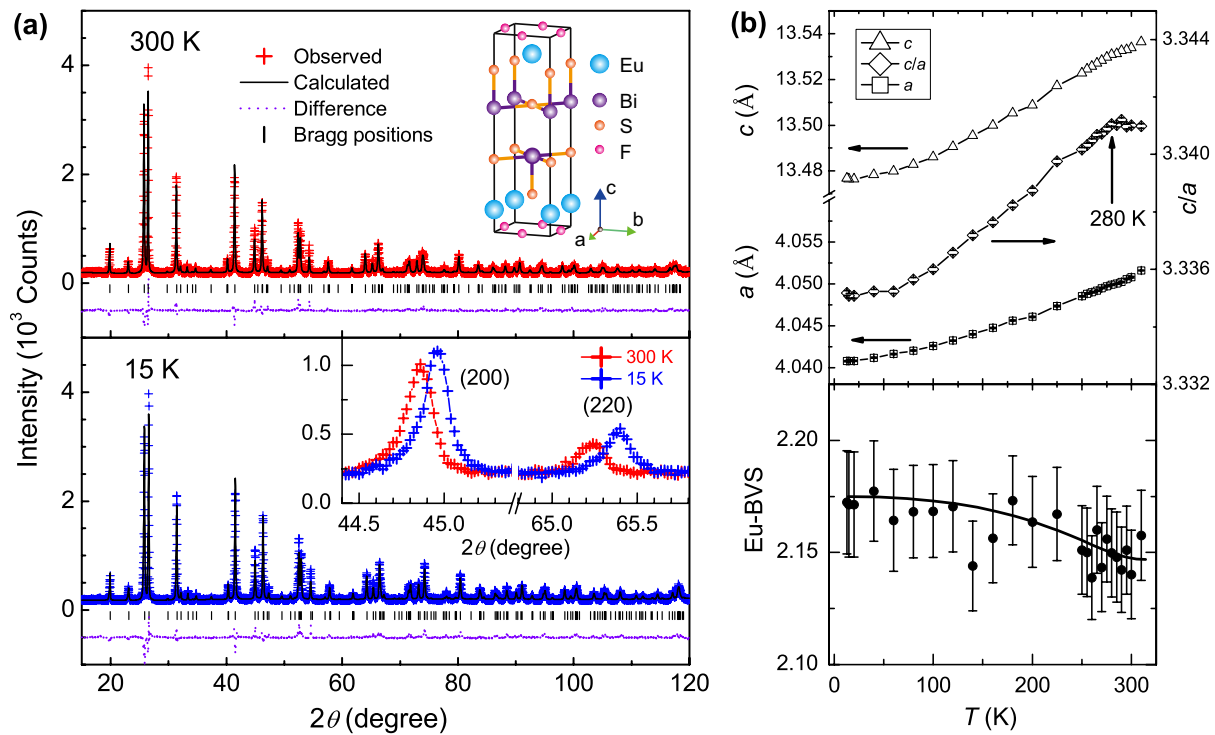}
\caption{\label{fig1}Crystal structure and its temperature dependence for EuBiS$_{2}$F. (a) Powder x-ray diffractions and their Rietveld refinement profiles at 300 K (top) and 15 K (bottom). Top inset: the crystal structure of EuBiS$_{2}$F. Bottom inset: an enlarged plot showing the (200) and (220) reflections. (b) Temperature dependence of lattice parameters $a$, $c$, $c/a$, and the bond valence sum of Eu (Eu-BVS). The solid line is a guide to the eye.}
\end{figure*}

EuBiS$_{2}$F crystallizes in the tetragonal CeBiS$_2$O-type
structure\cite{ceolin} with space group $P4/nmm$ (No. 129). The
crystal structure, as depicted in the top inset of Fig.~\ref{fig1}(a), can be viewed as an intergrowth of fluorite-like Eu$_2$F$_2$ block layers and
NaCl-like BiS$_2$ bilayers along the crystallographic $c$ axis. The
XRD patterns for the EuBiS$_{2}$F sample
were well reproduced using the crystal structure model by Rietveld
refinement\cite{rietan}. No obvious extra reflections appear,
indicating monophasic EuBiS$_{2}$F within the XRD detecting limit
($\sim$ 2 wt.\% in normal conditions).

The refined structural data were tabulated in Table S1 of the Supplemental Material (SM)\cite{SM}. The room-temperature lattice parameters $a$ [=4.0508(1)
{\AA}] and $c$ [=13.5338(3) {\AA}] are 0.7\% and 2.1\% smaller,
respectively, than the counterparts of SrBiS$_{2}$F [$a$=4.079(2)
{\AA} and $c$=13.814(5) {\AA}\cite{lei}]. As a consequence, the
$c/a$ ratio of EuBiS$_{2}$F [3.341(1)] is remarkably reduced,
compared with those ($\sim$ 3.39) of other CeBiS$_{2}$O-type parent
compounds. Since the $c/a$ value decreases upon electron doping for
the CeBiS$_{2}$O-related
systems\cite{mizuguchi,demura,whh,awana,jha,lyk}, one may speculate
that EuBiS$_{2}$F could be self electron doped (see previous
examples of self doping in Refs. \cite{cao,sun}) owing to possible
mixed valence of Eu. An alternative approach to verify this
speculation is to calculate the bond valence sum\cite{BVS} of Eu
(Eu-BVS) by the formulae $\sum
\text{exp}(\frac{R_{0}-d_{ij}}{0.37})$, where $R_{0}$ is empirical
parameters (2.04 {\AA} and 2.53 {\AA} for Eu$-$F and Eu$-$S bonds,
respectively\cite{BVS}) and $d_{ij}$ denotes the measured bond
distances between Eu and the \emph{nine} coordinating anions. The
resultant Eu-BVS values are 2.14(2) $-$ 2.18(2), weakly depending on
temperature (in comparison, the Sr-BVS value in SrBiS$_{2}$F is
calculated to be 2.05(4) at based on the
crystallographic data in Ref. \cite{lei}). This means that the
BiS$_2$ bilayers are indeed self doped, accounting in turn for the
anomalously small $c/a$ ratio in EuBiS$_{2}$F .

In order to detect the possible superstructured CDW phase
predicted\cite{wxg,yildirim}, we performed low-temperature XRD
measurements down to 13 K. No splitting for the (200) or (220) peak
was detectable, as seen in the bottom inset of Fig.~\ref{fig1}(a).
Also, the structural fittings using the superstructures of
$\sqrt{2}\times\sqrt{2}\times 1$\cite{wxg} (with space group $Cmma$)
or $\sqrt{2}\times2\sqrt{2}\times 1$\cite{yildirim} (with space
group $P22_{1}2$) could not give a better refinement. Although static long-range CDW order cannot be detected by the conventional XRD technique, considering the strong evidences for a CDW transition at 280 K (see the following sections), we suppose that a dynamic and/or short-range CDW ordering is still likely. Indeed, theoretical calculations\cite{wxg} indicate shallowness of the double-well potential with respect to the in-plane S(1) displacement, which leads to absence of static CDW order. Further experimental investigations with other techniques such as synchrotron XRD
and electron diffractions at low temperatures are expected to be
helpful to clarify this issue.

Figure~\ref{fig1}(b) show temperature dependence of the crystal
structural parameters and the Eu-BVS. The lattice parameters $a$ and
$c$ decrease monotonically with decreasing temperature. A subtle
anomaly around 280 K can be detected, and it is more evident in
$c/a$ ratio. As stated above, the decrease in $c/a$ means more
electron doping on the BiS$_2$ bilayers. Thus, the small decrease in $c/a$ below 280 K suggests further electron transfer from the Eu atoms to the
BiS$_2$ bilayers with decreasing temperature. Indeed, the Eu-BVS
value exhibits a detectable variations. At lower temperatures, both
$c/a$ and Eu-BVS saturate, indicating a stable mixed-valence state
for Eu.

\subsection{\label{subsec:level2}Electrical transport properties}

\begin{figure*}
\includegraphics[width=15cm]{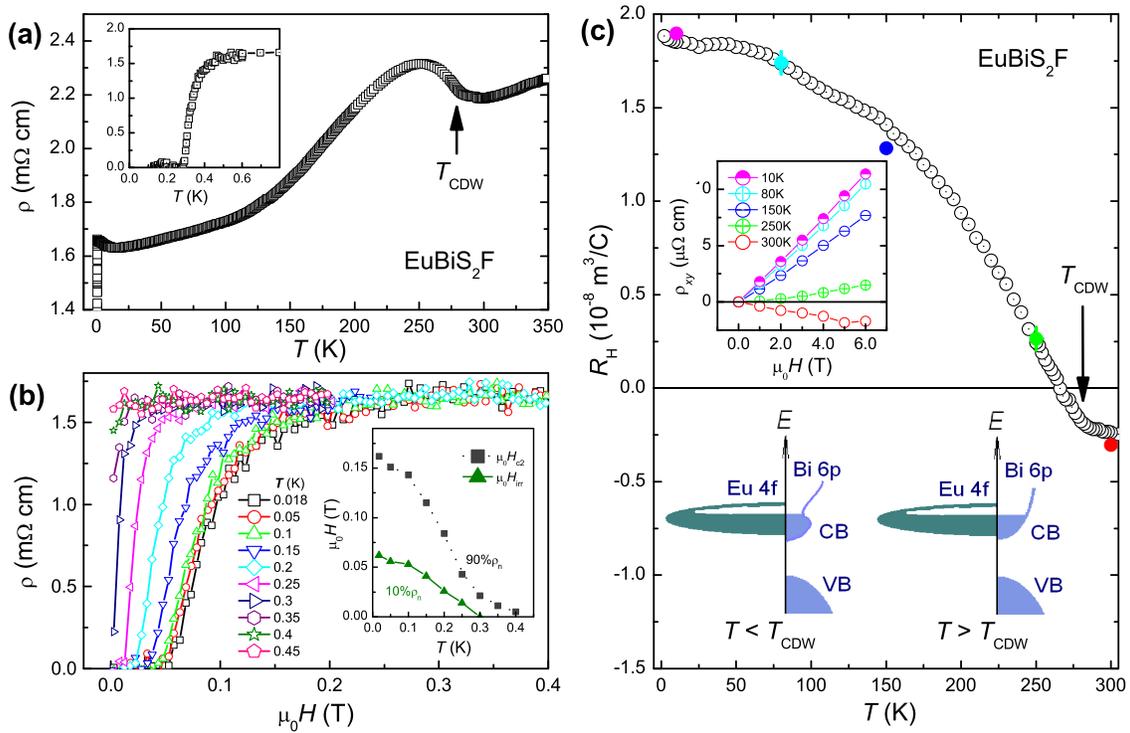}
\caption{\label{fig2}Electrical transport properties and
superconductivity in EuBiS$_{2}$F. (a) Temperature dependence of
resistivity showing a CDW-like anomaly at 280 K and
superconductivity at 0.3 K (inset). (b) The magnetoresistivity as a
function of magnetic field at low temperatures. The inset plots the
upper critical field, $H_{\text{c2}}$, by the criteria of
90\%$\rho_{n}$, where $\rho_n$ stands for the normal-state
resistivity. (c) Temperature dependence of Hall coefficient in which
a turning point at 280 K is seen. Upper inset: field dependence of
Hall resistivity at some fixed temperatures. Bottom inset: schematic
energy-band diagrams for $T<T_{\text{CDW}}$ (left) and
$T>T_{\text{CDW}}$ (right), respectively. CB (VB) denotes thy
conduction band (valence band) in the BiS$_2$ bilayers.}
\end{figure*}

Figure~\ref{fig2} shows the temperature dependence of resistivity,
$\rho(T)$, for the EuBiS$_{2}$F polycrystalline sample. Unlike the
parent compound SrBiS$_2$F that shows semiconducting
behaviour\cite{lyk,lei}, EuBiS$_{2}$F is virtually metallic, due to
the self doping effect. More surprisingly, the $\rho(T)$ curve exhibits a
broad hump below $\sim$ 280 K, resembling the CDW transitions in Q2D
systems like Cu$_x$TiSe$_2$\cite{morosan}. The resistivity anomaly
cannot be ascribed to the increase of the Eu valence, because the
latter would generate more electron carriers in the conducting
BiS$_2$ bilayers, which would lower (rather than raise) the
resistivity. Moreover, the hump is not even related to the Eu-4$f$
electrons, because we observed a similar hump, due to the CDW
instability, in an Eu-free sample
(Sr$_{0.7}$Ca$_{0.3}$)$_{0.75}$La$_{0.25}$BiS$_{2}$F (see Fig. S1 in the SM\cite{SM}). Therefore, this $\rho(T)$ hump is interpreted
by a gap (probably a pseudogap) opening at Fermi level
($E_{\text{F}}$) because of the formation of dynamic/short-range CDW
below 280 K. Here we note that no obvious nonlinear $I-V$
relations associated with the sliding of CDW was observed down to 2 K.

At lower temperatures, a superconducting transition takes place with
a zero-resistance temperature of 0.3 K at zero field [see the inset
of Fig.~\ref{fig2}(a)]. The superconducting transition temperature
$T_\text{c}$ is reduced by a factor of 10, compared with the
BiS$_2$-based superconductors synthesized under ambient
pressure\cite{demura,whh,awana,jha,lyk}. This could be due to the
formation of CDW which loses partial FSs, relatively low electron
doping, and pair breaking by the Eu magnetic moment. The
superconducting transition was also demonstrated by the
magnetoresistivity measurement at fixed temperatures, as shown in
Fig.~\ref{fig2}(b). The upper critical fields ($H_{\text{c2}}$) were
determined using the criteria of 90\%$\rho_{n}$, where $\rho_n$
stands for the normal-state resistivity. The temperature dependence
of $H_{\text{c2}}$ shows positive curvature below $T_\text{c}$,
possibly due to large anisotropy in $H_{\text{c2}}$\cite{Hc2} (note
that the sample is polycrystals). The $\mu_{0}H_{\text{c2}}$ value
at 0.018 K is 0.16 T, which is obviously smaller than the Pauli
limiting field, $\mu_{0}H_{\text{P}}$=1.84 $T_c$ $\approx$ 0.4 T.

To further understand the anomaly around 280 K in $\rho(T)$,
we measured the temperature dependence of Hall coefficient
($R_{\text{H}}$). As shown in Fig.~\ref{fig2}(c), at room
temperature, $R_{\text{H}}$ is negative, indicating dominant
electron transport. If assuming single band scenario, the carrier
density is estimated to be
$n=1/e|R_{\text{H}}|=(2.1\pm0.3)\times10^{27}$ m$^{-3}$, equivalent
to a Hall number of $V_{\text{cell}}/(2e|R_{\text{H}}|)=0.24\pm0.03$
electrons per formula unit (fu). This electron density corresponds
to an Eu valence of +2.24(3), if assuming that all the electron
carriers in the conduction band (CB) come from the Eu-4$f$ orbitals
[see the schematic energy-band diagrams in Fig.~\ref{fig2}(c)]. Below 280 K, $R_\text{H}$ increases steeply,
and then it changes the sign at lower temperatures. This peculiar
$R_\text{H}(T)$ behaviour strongly suggests FS reconstructions owing
to a CDW transition. In Sr$_{1-x}$La$_{x}$BiS$_2$F system, as a
comparison, the $R_\text{H}(T)$ is either positive (for $x\leq$0.45)
or negative (for $x\geq$0.5) \cite{lyk,sakai,lyk2}. We speculate
that a pseudogap opens below $T_{\text{CDW}}$, which could lower the
$E_{\text{F}}$ a little, as shown in the left diagram in Fig.~\ref{fig2}(c). As a result, more transferred electrons are expected,
accounting for the small increase of Eu valence below 280 K.

\subsection{\label{subsec:level3}Magnetic properties}
The magnetic properties Eu$^{2+}$ and Eu$^{3+}$ ions are very
different because of different electron filling on the 4$f$ orbitals
(4$f^7$ and 4$f^6$, respectively). The ground state of the former is
$^8S_{7/2}$ with an effective local-moment of $g\sqrt{S(S+1)}=$7.94
($\mu_{\text{B}}$). In contrast, the ground state of the latter is
$^{7}F_{0}$, which has zero magnetic moment. Nevertheless, the
excited states $^7F_{J}$ ($J$=1, 2, ..., 6), due to the spin-orbit
interaction $\lambda \textbf{L$\cdot$S}$, give rise to appreciable
Van Vleck paramagnetic susceptibility
($\chi_{\text{vv}}$)\cite{vanvleck}. Here the coupling constant
$\lambda$ also measures the energy of the first excited state. The
$\lambda$ value is 480 K for free Eu$^{3+}$ ions\cite{vanvleck}, and
it has a small change in solids, e.g., $\lambda$=471 K for EuBO$_3$
and $\lambda$=490 K for EuF$_3$\cite{nagata}. Consequently, the
$\chi_{\text{vv}}(T)$ is featured by a temperature-independent
plateau in the low-temperature regime (say, $T\leq$ 100 K), and
Curie-like paramagnetism with an effective magneton number of 3.4
per Eu$^{3+}$ for high-temperature region ($T>$ 200
K)\cite{nagata}. In EuBiS$_2$F system, therefore, we can correctly
analyze the temperature dependence of magnetic susceptibility,
$\chi(T)$, with an extended Curie-Weiss law,
\begin{equation}
\label{eq1}
\chi=\chi_{0}+C/(T+\theta_{\text{N}}).
\end{equation}
In the low-temperature limit, the first term $\chi_{0}$ includes
$\chi_{\text{vv}}$, in addition to Pauli paramagnetism (and Landau
diamagnetism) of conduction electrons and Langevin diamagnetism from
the core-shell electrons of all the constituent elements. At the
high-temperature side, $\chi_{\text{vv}}$ is included to the second
term, where $C$ denotes Curie constant and $\theta_{\text{N}}$ is
termed as paramagnetic Neel temperature. One may obtain the
effective moment by the formulae
$\mu_{\text{eff}}=\sqrt{3k_{\text{B}}C/N_{\text{A}}}$, where
$k_{\text{B}}$ and $N_{\text{A}}$ denote Boltzmann and Avogadro
constants, respectively.

\begin{figure}
\includegraphics[width=8.5cm]{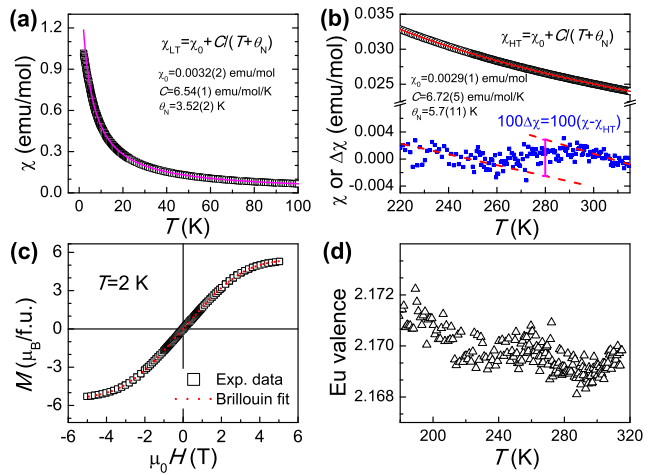}
\caption{\label{fig3} Magnetic properties of EuBiS$_2$F. Temperature
dependence of the dc magnetic susceptibility [(a): $T\leq$ 100 K;
(b): $T\geq$ 220 K]. The solid lines are the fitted curves. The
difference (multiplied by 100) between the experimental data and the
fitted ones is shown in the lower part of (b). (c) Field dependence
of magnetization at 2 K. The dotted line is a Brillouin fit. (d) The
Eu valence estimated by the magnetic susceptibility. See details in
the text.}
\end{figure}

Figure~\ref{fig3}(a) shows the temperature dependence of magnetic
susceptibility, $\chi(T)$, for the EuBiS$_2$F sample at $T\leq$ 100
K. No magnetic transition is evident down to 2 K. The $\chi(T)$ data
can be well fitted by Eq.~(\ref{eq1}), except for minor deviations
below 10 K. The fitted effective paramagnetic moment is 7.2
$\mu_{\text{B}}$ fu$^{-1}$, which means that the concentration of
Eu$^{2+}$ is 83\% (hence the Eu valence is +2.17, fully consistent
with the Eu-BVS value above). The Eu$^{3+}$ ions (with a population
of 17\%) should give considerable contribution to $\chi_0$. Indeed,
the fitted $\chi_0$ value is as large as 0.0032(2) emu mol$^{-1}$.
By referring to the Van Vleck susceptibility of EuF$_3$ ($\sim$
0.006 emu mol$^{-1}$ below 100 K) in which the Eu valence is totally
3+\cite{nagata}, the $\chi_{\text{vv}}$ value of EuBiS$_2$F is then
estimated to be 0.001 emu mol$^{-1}$ below 100 K. Since the Langevin
diamagnetic susceptibility is negligibly small (about
$-1.5\times10^{-4}$ emu mol$^{-1}$)\cite{core} within the fitting
errors, then, the Pauli susceptibility can be roughly estimated to
be $\sim$ 0.002 emu mol$^{-1}$. This unusually large value of Pauli
susceptibility suggests substantial hybridizations between Eu-4$f$
and CB (see band structure calculations in Section~\ref{subsec:level6}).

Figure~\ref{fig3}(b) shows the high-temperature range of $\chi(T)$,
which was also fitted with Eq.~(\ref{eq1}). The fitting gives smaller
$\chi_0$ but larger $C$ values, because $\chi_{\text{vv}}$ is now
included in the second term of Eq.~(\ref{eq1}). By subtraction of the
fitted curve from the experimental data, the residual susceptibility
$\Delta\chi$ shows a hump-like anomaly where the magnetic susceptibility
tends to drop [by $(5\pm2)\times10^{-5}$ emu mol$^{-1}$] at around 280 K. At first sight, it seems to be related to the change in Eu valence. However, similar hump in $\chi$ was also observed in an Eu-free analogous sample (Sr$_{0.7}$Ca$_{0.3}$)$_{0.75}$La$_{0.25}$BiS$_{2}$F (see Fig. S2 in the SM\cite{SM}). Thus we
speculate that the 'drop' of $\chi$ at 280 K is mainly resulted from the
loss of Pauli paramagnetic susceptibility when a (pseudo)gap opens
at the CDW transition. Here we note that the decrease of
$N(E_{\text{F}})$ in the CDW phase for LaBiS$_2$O$_{0.5}$F$_{0.5}$
calculated\cite{yildirim} is just equivalent to the loss of Pauli
magnetic susceptibility.

The possible variation in Eu valence can be analyzed as follows. If
$P_{3+}$ denotes the concentration of Eu$^{3+}$ [so that
the fraction of Eu$^{2+}$ is ($1-P_{3+}$)], one may calculate $P_{3+}$
by the relation,
\begin{equation}
\chi=\chi_{0}+P_{3+}\frac{C_{3+}}{T}+(1-P_{3+})\frac{C_{2+}}{T+\theta_{\text{N}}}.
\end{equation}
With $C_{2+}$=7.875 emu mol$^{-1}$ K$^{-1}$ and $C_{3+}$=1.45 emu
mol$^{-1}$ K$^{-1}$\cite{nagata}, the Eu valence was
obtained as plotted in Fig.~\ref{fig3}(d). One sees that the Eu valence does
not change so much around 280 K, but it tends to increase rapidly below 220
K, which is quantitatively consistent with the Eu-BVS
values shown in Fig.~\ref{fig1}(b).

One may also obtain the information of Eu valence from the
field-dependent magnetization at 2 K [Fig.~\ref{fig3}(c)] which
shows a saturation at high magnetic fields. The saturation
magnetization is significantly
smaller than the expected value ($gJ$=7.0 $\mu_{\text{B}}$
fu$^{-1}$) for Eu$^{2+}$ions only. By fitting the $M(H)$ data using
a Brillouin function with consideration of Weiss molecular field
$B_{\text{mf}}=\xi M$, the saturation magnetization is determined to
be 5.58 $\mu_{\text{B}}$ fu$^{-1}$, corresponding to the Eu valence
of +2.20, consistent with the conclusion from the Curie-Weiss
fitting above. The fitted parameter $\xi$ is a negative value
($-$0.40), reflecting dominant antiferromagnetic interactions among
the Eu localized moments, also agreeing with the positive value of
$\theta_{\text{N}}$ in Eq.~(\ref{eq1}).

\subsection{\label{subsec:level4}M\"{o}ssbauer spectroscopy}
M\"{o}ssbauer spectroscopy (MS) is a powerful technique to study the
Eu valence. The isomer shift $S$ of the nuclei of Eu$^{2+}$ and
Eu$^{3+}$ ions falls in two nonoverlapping ranges: $S_{2+}$=
$-7.7\sim-13.5$ mm s$^{-1}$ and $S_{3+}$= $-0.01\sim+2.6$ mm
s$^{-1}$ (relative to Eu$_2$O$_3$). Thus, MS easily identifies the
valence state of Eu and, in the case of inhomogeneous mixed valence,
the concentration of each Eu species can be determined by the
relative absorption intensity. For the fast valence fluctuation (VF)
scenario, the mean Eu valence can also be quantitatively evaluated
by the change in $S$\cite{EuCu2Si2}. MS may also supply information
on quadrupole and magnetic interactions.

\begin{figure}
\includegraphics[width=8.5cm]{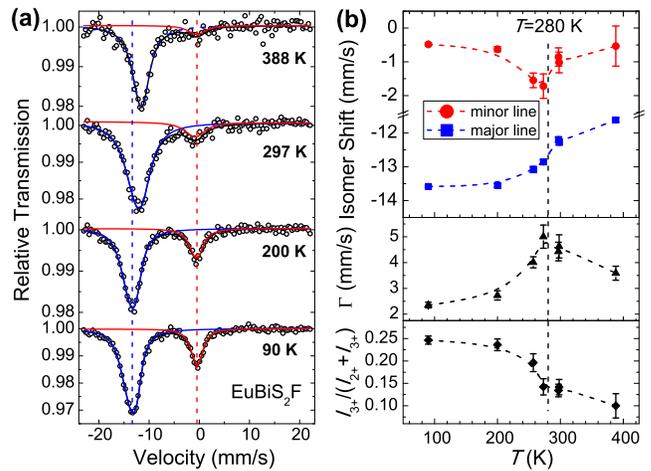}
\caption{\label{fig4}M\"{o}ssbauer result on EuBiS$_2$F. (a)
$^{151}$Eu M$\ddot{\text{o}}$ssbauer spectra of EuBiS$_2$F at some
typical temperatures. (b) Temperature dependence of the isomer
shifts (top), absorption line width (middle), relative intensity
of the minor line (bottom) from the data fitting of the $^{151}$Eu
M\"{o}ssbauer spectra. The vertical dashed lines at $-$13.5 and
$-$0.5 mm/s (a) and at 280 K (b) are guides to the eye.}
\end{figure}

Figure~\ref{fig4}(a) shows the $^{151}$Eu M\"{o}ssbauer spectra at
some typical temperatures of 90, 200, 297 and 388 K. At low
temperatures (90 and 200 K), two absorption lines (or peaks) appear
at $-$13.5 and $-$0.5 mm s$^{-1}$, which are obviously identified to
the M\"{o}ssbauer resonance absorptions of Eu$^{2+}$ and Eu$^{3+}$
nuclei, respectively. Since there is only one equivalent site in the
crystal structure even at low temperatures, the two separate lines
indicate slow Eu VFs (or even static charge ordering of Eu$^{2+}$
and Eu$^{3+}$) with the time scale of $\tau_{\text{vf}}>10^{-8}$ s
(note that the probing time of MS is about 10$^{-9}$ s). The
intensity of the minor line is about 1/3 of the major one,
therefore, the mean Eu valence is $\sim$ +2.25, basically consistent
with the above conclusion drawn from the crystal structure and
magnetic data. As the temperature is increased to 297 K, the
intensity of the Eu$^{3+}$ line decreases. Simultaneously, the two
absorption lines get closer, and the line width is abnormally large
(up to 5 mm s$^{-1}$). This fact suggests Eu VFs with the
characteristic $\tau_{\text{vf}}$ close to 10$^{-9}$ s, associated
with the electron hopping between Eu$^{2+}$ and Eu$^{3+}$. At 388 K,
the Eu$^{3+}$ line almost vanishes because the VFs are much faster.
The isomer shift is then formulated by
$S_{\text{vf}}=(1-P_{3+})S_{2+}+P_{3+}S_{3+}$. With
$S_{\text{vf}}=-11.6$ mm s$^{-1}$, $S_{2+}=-$13.5 mm s$^{-1}$ and
$S_{3+}=0 $ mm s$^{-1}$, the Eu valence at 388 K can be estimated to
be +2.14(2).

By the data fitting using the exact shape of the emission spectrum
of $^{151}$Sm$_2$O$_3$ and considering quadrupole interactions, we
were able to obtain the refined $S$, the full linewidth at half
maximum ($\Gamma$) and the peak intensities ($I$), which are plotted
respectively in Fig.~\ref{fig4}(b). All these MS parameters point to
a transition at 280 K. The most prominent feature is that the
$\Gamma$ value is peaked at 280 K. The sharp decrease in $\Gamma$
below 280 K suggests that the Eu VFs slow down, probably in
connection with the dynamic CDW. The isomer shifts and the relative
intensity of the minor line respectively reflect fast VFs with
$\tau_{\text{vf}}<10^{-9}$ s and slow VFs with
$\tau_{\text{vf}}>10^{-9}$ s. At 273 K, for example, one may roughly
estimated that 5(1)\% of the Eu ions are in Eu$^{3+}$ state with
fast VFs, and 14(2)\% of the Eu ions are in Eu$^{3+}$ state with
slow VFs. So, the overall Eu valence is +2.19(3) at 273 K.

\subsection{\label{subsec:level5}Specific heat}
\begin{figure}
\includegraphics[width=8cm]{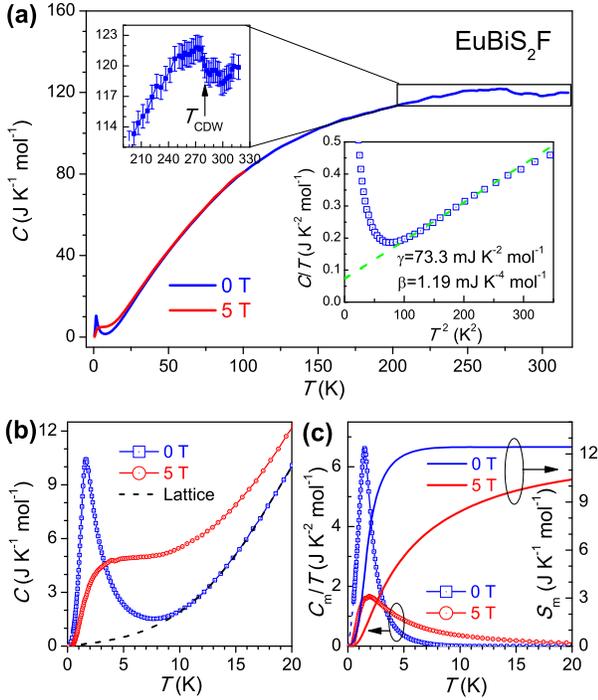}
\caption{\label{fig5}Specific heat capcity for EuBiS$_{2}$F. (a)
Temperature dependence of specific heat capacity, $C(T)$, from 0.5
to 320 K. Upper left inset: an enlarged plot of the $C(T)$ showing
an anomaly at 280 K. Lower right inset: plot of $C/T$ vs. $T^2$ at
low-temperature region (the dashed line obeys $C/T=\gamma+\beta
T^2$). (b) An enlarged plot of $C(T)$ from 0.5 to 20 K. The dashed
line is a polynomial fit (see details in the text), representing the
lattice contribution. (c) $C_{\text{m}}/T$ (where $C_{\text{m}}$
denotes magnetic contribution to the specific heat) and magnetic
entropy $S_\text{m}$ (left axis) as functions of temperature.}
\end{figure}

Heat capacity of a solid may supply important information not only
for phase transition but also for electronic and magnetic states.
Fig.~\ref{fig5}(a) shows the temperature dependence of specific
heat, $C(T)$, of EuBiS$_{2}$F. The $C(T)$ data tend to saturate to
120 J K$^{-1}$ mol$^{-1}$ at room temperature, consistent with the
high-$T$ limit for lattice specific heat (i.e., the Dulong-Petit
value 3$NR$=15$R$=124.7 J K$^{-1}$ mol$^{-1}$, where $N$ counts the
number of elements per fu, $R$ is the gas constant). One sees an
anomaly around 280 K, a signal of second-order transition, further
supporting a CDW-like transition.

In EuBiS$_{2}$F, the specific heat is contributed due to several
different origins including crystalline lattice ($C_{\text{lat}}$),
conduction electron ($C_{\text{el}}$) and Eu magnetism
($C_{\text{m}}$). One may correctly separate them out by considering different contribution
weights in different temperature regions. Since no magnetic ordering
takes place above 2 K, and the magnetic susceptibility well follows
Curie-Weiss law above 10 K, $C_{\text{m}}$ is then expected to be very small above 10 K at zero magnetic field. Therefore, we make use
of a conventional approach, in which the low-$T$ lattice
contribution is taken as $\beta T^3$, to extract $C_{\text{el}}$.
The plot of $C/T$ vs $T^2$ in the inset of Fig.~\ref{fig5}(a) gives
$\gamma$=73.3 mJ K$^{-2}$ mol$^{-1}$ and $\beta$=1.19 mJ K$^{-4}$
mol$^{-1}$. The resultant Debye temperature,
$\theta_{\text{D}}=[(12/5)NR\pi^{4}/\beta]^{1/3}$=201 K, is
reasonably in between with those of LaBiS$_2$O$_{0.5}$F$_{0.5}$ (221
K) and YbBiS$_2$O$_{0.5}$F$_{0.5}$ (186 K)\cite{yazici}, which vice
versa guarantee the reliability of the Sommerfeld parameter.
Notably, the fitted $\gamma$ value is over 50 times of that of
Sr$_{0.5}$La$_{0.5}$BiS$_2$F (1.42 mJ K$^{-2}$
mol$^{-1}$)\cite{lyk}. Its corresponding $N(E_{\text{F}})$
[$=3\gamma /(\pi k_{\text{B}})^{2}$] is as large as 30 eV$^{-1}$
fu$^{-1}$, which is about 25 times of the bare density of states of
LaBiS$_2$O$_{0.5}$F$_{0.5}$  (1.22 eV$^{-1}$ fu$^{-1}$)\cite{li}.
The greatly enhanced $\gamma$ is related to the unusually large
Pauli magnetic susceptibility above, mainly originating from the
hybridization between conduction electrons and the Eu 4$f$ electrons
(see band structure calculations in Section~\ref{subsec:level6}).

Figure~\ref{fig5}(b) zooms in the $C(T)$ data below 20 K. At zero
magnetic field, a peak appears at 1.6 K with a high maximum up to
10.41 J K$^{-1}$ mol$^{-1}$. Under a magnetic field of 5 T, the peak
is suppressed, forming a broad hump centered at about 3.5 K. Since the
ground state of Eu$^{2+}$ has zero orbital angular momentum,
Schottky-like contribution is not expected. Hence the peak should be of
magnetic origin. It is noted that, on the right side of the peak,
there is no specific jump (or a divergence), as opposed to an ordinary
long-range magnetic ordering. This lets us consider that the
specific anomaly comes from freezing of non-ordered Eu$^{2+}$ spins,
i.e., a spin glass transition. Similar observation was reported in
(Eu,Sr)S\cite{meschede} and EuCu$_2$Si$_2$\cite{wang}. To extract $C_{\text{m}}$ below 10 K more accurately, the $C(T)$ data from 10 K to 20 K
was fitted by a polynomial with odd-power terms, $C\approx
C_{\text{el}}+C_{\text{lat}}=A_{1}T+A_{3}T^{3}+A_{5}T^{5}+A_{7}T^{7}$
(the resultant $A_{1}$ and $A_{3}$ agree well with the above
$\gamma$ and $\beta$ values). Then the magnetic contribution
$C_{\text{m}}$ below 10 K was obtained by removing the contributions
of $C_{\text{lat}}$ and $C_{\text{el}}$. Consequently, the magnetic
entropy can be calculated by $S_{\text{m}}=\int_{0}^{T}
(C_{\text{m}}/T)dT$, as shown in Fig.~\ref{fig5}(c), which is 12.4 J
K$^{-1}$ mol$^{-1}$ at 20 K under zero field. Furthermore, nearly
the same $S_{\text{m}}$ value can be achieved under 5 T for
integrating up to 60 K. The released magnetic entropy equals to 72\%
of $R$ln($2S+1$) ($S$=7/2 for Eu$^{2+}$). Considered omission of magnetic entropy
above 20 K, the resultant $S_{\text{m}}$ value actually gives an upper
limit of the Eu valence of +2.28 below 20 K.

\subsection{\label{subsec:level6}Band structure calculations}
\begin{figure*}
\includegraphics[width=15cm]{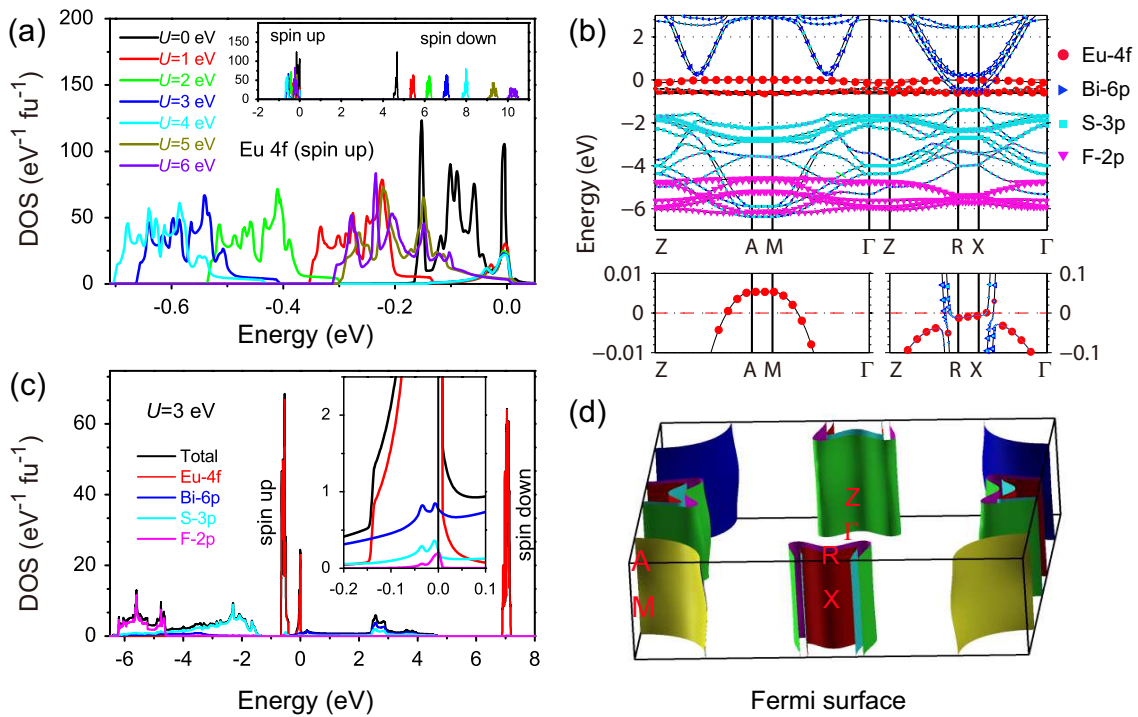}
\caption{\label{fig6}Band calculations for EuBiS$_{2}$F using
LSDA+$U$ method. (a) Variations of Eu-4$f$ bands with different $U$
values. DOS denotes to density of states. (b) Calculated band
structure of EuBiS$_{2}$F with $U$=3 eV. The contributions of the
relevant orbital states are distinguished by different colors. The
lower panels zoom in the band dispersions crossing the Fermi level
($E_\text{F}$=0). (c) Total and projected DOS with $U$=3 eV. Inset:
an enlarged plot near $E_\text{F}$. (d) Fermi surfaces of
EuBiS$_{2}$F derived from the band structure in (b).}
\end{figure*}

To interpret above experimental results, we performed a
first-principles calculation using local spin density approximation
with consideration of on-site Coulomb interaction (LSDA+$U$),
particularly paying attention to the Eu-4$f$ electronic states (the
issue of CDW instability was well documented by
Yildirim\cite{yildirim}). We first investigated the influence of the
parameter $U$. Fig.~\ref{fig6}(a) shows the variations of Eu-4$f$
bands of EuBiS$_{2}$F with $U$=0, 1, ..., 6 eV. In all the cases,
there is a large gap between occupied and unoccupied levels, as
shown in the inset. Another prominent feature is that the highest
occupied band (HOB) locates around $E_\text{F}$, regardless of the
different $U$ values. This fact suggests that the Eu-4$f$ frontier
occupied level in the HOB is pinned by the chemical potential of the
conduction bands. The filled states in HOB actually represents
Eu$^{2+}$, whilst the empty states in HOB, i.e., the 4$f$ holes,
correspond to Eu$^{3+}$.

Figure~\ref{fig6}(b) shows the calculated band structure of
EuBiS$_{2}$F with a realistic value of $U$=3 eV. There are seven
flat bands near $E_\text{F}$, all coming from the Eu-4$f$ orbitals.
The HOB crosses $E_\text{F}$, and hybridizes with the Bi-6$p$ bands
(see the zoom-in plot at the bottom). Furthermore, this HOB donates
electrons to the Bi-6$p$ bands, and leaves hole pockets around the
\textbf{M} point [see also the FS in Fig.~\ref{fig6}(d)]. Consequently,
although undoped, the CB of EuBiS$_{2}$F is
filled with the transferred electrons. Due to the Q2D
structure, cylindric-like FSs are presented [Fig.~\ref{fig6}(d)].
Except for the 4$f$-hole pockets, two electron-type FS sheets appear
around the \textbf{X} point, similar to the case of
LaBiS$_2$O$_{1-x}$F$_{x}$ with $x$=0.25\cite{kuroki}. This 2D-like
FS sheets have considerable nesting areas for developing CDW
instability.

Figure~\ref{fig6}(c) shows the calculated electronic DOS of
EuBiS$_{2}$F. Obviously, the sharp peaks come from the Eu-4$f$
orbitals. They are basically divided into two groups, which are
about 7 eV apart (corresponding to an effective Hubbard $U\sim$ 7
eV). The frontier HOB nearby $E_\text{F}$ is mainly contributed from
the $f_{xz^{2}}$ wave function. The DOS at $E_\text{F}$,
$N(E_\text{F})$, mainly comprises of Bi-6$p$ (0.75$\times$2
eV$^{-1}$ fu$^{-1}$) and Eu-4$f$ (21.6 eV$^{-1}$ fu$^{-1}$). The
total $N(E_\text{F})$ is about 20 times of that of
LaBiS$_2$O$_{0.5}$F$_{0.5}$\cite{li}, well accounting for the
enhancement of electronic specific-heat coefficient as well as Pauli
susceptibility.

\section{\label{sec:level4}Concluding remarks}

We have demonstrated a series of signatures for a CDW
transition in EuBiS$_{2}$F, albeit no static long-range superlattice
order was detected by XRD. They include a clear kink in the $c/a$
ratio, a resistivity hump, a kink in Hall coefficient, a subtle magnetic susceptibility drop, and a specific-heat hump, all at
$T_{\text{CDW}}\sim$ 280 K. The slowing down of Eu VFs below
$T_{\text{CDW}}$ suggests a dynamic CDW ordering.
Such a dynamic CDW is supported by the theoretical calculations
which indicate dynamic in-plane displacements of S(1) owing to the
shallowness of the double-well potential\cite{wxg}. Very recently,
'checkerboard stripe' electronic state was observed on the cleaved
surface of NdO$_{0.7}$F$_{0.3}$BiS$_2$ single crystals\cite{stm}.
Although the nanoscale electronic inhomogeneity was considered to be
due to the atomic defects on the cleaved surface, it could be in
some relations to the CDW instability.

\begin{table*}
\caption{\label{tab1}Summary of the Eu valence in
EuBiS$_{2}$F determined via various methods in different temperature ($T$)
ranges. Eu-BVS refers to bond valence sum\cite{BVS} of the Eu ions.
The number in parentheses represents the measurement uncertainty for the last digit.}
\begin{ruledtabular}
\begin{tabular}{lcccccr}
Methods & Eu-BVS & Magnetization & M\"{o}ssbauer & Heat capacity & Fermi surface\\
\colrule
Eu valence &2.14(2)$-$2.18(2)&2.17(2)$-$2.20(1)&2.24(2); 2.19(3); 2.14(2)&$<$ 2.28&2.25(5)\\
$T$  (K) &310$-$13 &300$-$2 &$\leq$ 200; 273; 388 &0.5$-$20&N.A. \\
\end{tabular}
\end{ruledtabular}
\end{table*}

As is known, CDW mostly originates from the FS
nesting\cite{review,cdw}. Since the shape and the size of the
FS sheets are predominantly decided by the electron filling in the CB
of BiS$_{2}$-based materials\cite{kuroki}, the electron doping level
should be crucial for the occurrence of CDW. Our finding in
EuBiS$_{2}$F implies that the CDW instability is optimized at around
$x\sim$ 0.2. To verify this point, we synthesized samples
of (Sr,Ca)$_{0.75}$La$_{0.25}$BiS$_{2}$F, in which the electron
doping was fixed to $x$=0.25. We indeed observed a similar CDW anomaly
in this designed system (see Figs. S1 and S2 in SM\cite{SM}).

It is the mixed valence of Eu that considerable amount of electron carriers are transferred into the CB, which induces SC as well as CDW. Table~\ref{tab1} summarizes the Eu valence in EuBiS$_{2}$F determined via
various methods at different temperature ranges. Basically, the Eu
valence is about +2.2, nearly independent of temperature down to 2
K. This result is very unusual, since the Eu valence mostly increases remarkably with decreasing temperature, e.g., in the systems of
Eu$M_2$Si$_2$ ($M$=Cu\cite{EuCu2Si2}, Pd\cite{EuPd2Si2},
Ir\cite{EuIr2Si2,steglich}) and
EuNi$_2$P$_2$\cite{nagarajan,steglich} where Eu VFs were present. As
for the crystal structure of EuBiS$_{2}$F, there is only one
crystallographic site for the Eu ions. Therefore, the mixed valence of
Eu means existence of VFs. Indeed, according to the M\"{o}ssbauer
results above, VFs are dominant above $T_{\text{CDW}}$. Around
$T_{\text{CDW}}$, however, the VF frequency decreases rapidly. At
temperature far below $T_{\text{CDW}}$, the VFs of Eu$^{2+}$ and
Eu$^{3+}$ are slower than 10$^{8}$ Hz. These observations suggest that Eu
should have unequivalent Eu sites in short time scale, which could be realized in the proposed CDW phase with distortions of BiS layers\cite{yildirim}. Thus, the Eu valence is 'pinned' by the formation of CDW, resulting in the unique
temperature-independent mixed-valence. In turn, the Eu VFs (no matter how slow they are) should be in favor of an unusual dynamic CDW state.

The emergence of SC under the CDW transition suggests that
EuBiS$_{2}$F is also a CDW superconductor, like the well known NbSe$_2$\cite{review,cdw,wilson}. The formation of CDW generally
loses a portion of FSs, which leads to a decrease of $T_{\text{c}}$. In this sense, CDW competes with SC, as usual. However, an anharmonic model calculation\cite{wxg} shows that the CDW instability is also essential for the SC, suggesting more profound relationship between SC and CDW.

Finally we would like to emphasize that the occurrence of SC at the (self)
electron doping level corresponding to $x\sim$ 0.2 is also surprising. Previous reports\cite{sakai,lyk2} indicate that, in an
analogous system of Sr$_{1-x}$La$_{x}$BiS$_2$F, SC appears for $x>$
0.3, below which the samples show \emph{insulating} behaviour, and the
optimal doping is at $x\sim$ 0.5. According to the band structure
calculations\cite{kuroki,wxg,li,yildirim}, the FSs undergo a Lifshitz transition,
characterized by a change in FS topology, with electron doping. The two doping levels, $x\sim$ 0.2 and $x\sim$ 0.5, have very different FS topologies. Therefore, the superconducting state in EuBiS$_{2}$F might be
different from those of other BiS$_2$-based materials. It was predicted that\cite{wqh}, below the Lifshitz filling level (like the case of EuBiS$_{2}$F), the superconducting state is weak topological due to strong spin-orbit coupling. Further investigations are called for to verify this prediction.

\begin{acknowledgments}
We would like to thank F. Steglich, Q. M. Si, F. C. Zhang, J. H.
Dai, H. Q. Yuan, Q. H. Wang, C. Cao and Z. Ren for helpful discussions. This
work was supported by the National Basic Research Program of China
(under grants 2011CBA00103 and 2010CB923003), the National Science
Foundation of China (under grants 11190023), and the Fundamental
Research Funds for the Central Universities of China (Grant No.
2013FZA3003).
\end{acknowledgments}

%

\end{document}